\documentclass[12pt]{article}
\usepackage[english]{babel}
\usepackage{epsfig}
\begin{document}
\large
\par
{\bf Is neutrino produced in standard weak interactions a Dirac or
Majorana particle?}
\\

\begin{center}
\par
\vspace{0.3cm} Beshtoev Kh. M. (beshtoev@cv.jinr.ru)
\par
\vspace{0.3cm} Joint Institute for Nuclear Research, Joliot Curie
6, 141980 Dubna, Moscow region, Russia.
\end{center}
\vspace{0.3cm}

{\bf Abstract}

\par
This work considers the following problem: what type (Dirac or
Majorana) of neutrinos is produced in standard weak interactions?
It is concluded that only Dirac neutrinos but not Majorana
neutrinos can be produced in these interactions. It means that
this neutrino will be produced in another type of interaction.
Namely, Majorana neutrino will be produced in the interaction
which differentiates spin projections but cannot differentiate
neutrino (particle) from antineutrino (antiparticle). This
interaction has not been discovered yet. Therefore experiments
with very high precision are important to detect the neutrinoless
double decay. \\

\par
\noindent PACS: 14.60.Pq; 14.60.Lm

\section{Introduction}

The standard model \cite{1} is composed by using Dirac fermions
\cite{1dirac}. At present time violation of the electron
($l_{\nu_e}$), muon ($l_{\nu_\mu}$), tau ($l_{\nu_\tau}$) numbers
have been detected, then only the common lepton number is
conserved but not every lepton number individually. Then there is
a question: Does full violation of lepton numbers take place? In
this case the Majorana neutrino \cite{2maj} can be realized where
the conservation lepton number does not appear. Realization of the
Majorana neutrino can be fulfilled in two ways:
\par
1. To suppose that a priori neutrino is a Majorana particle, i.
e., to work in the framework of the standard model where it is
presumed that neutrinos are Majorana particles.
\par
2. To search for a condition (interaction) when Majorana neutrino
is produced.
\par
Almost in all works \cite{4} $\div$ \cite{9} where the Majorana
neutrino is investigated the first approach is used and there is a
conclusion that there is no possibility to differ is neutrino a
Majorana or Dirac particle, except the case of neutrinoless double
beta decay. In work \cite{10} it has been shown that for chiral
invariance of the standard weak interactions the neutrinoless
double beta decay cannot be realized.
\par
In this work we will use the second approach. In means that it is
necessary to find when and how (or in which interactions) the
Majorana neutrino can be produced. This approach is based on the
fundamental physical principle that in every interaction the
particles are produced in eigenstates. Then our aim is to find in
which type of interactions the Majorana neutrino can be produced
as eigenstate.
\par
Now come to a consideration of properties of Dirac and Majorana
particles.

\section{Dirac neutrino}

\par
The equation for a particle with spin $\frac{1}{2}$ was first
formulated by Dirac~\cite{1dirac} in 1928. Afterwards it turned
out that this representation was adequate to describe neutral and
charged fermions, i.\,e., fermions are Dirac particles. Dirac
equation for free fermion (spinor with spin $\frac{1}{2}$) has the
following form:
$$
(\gamma^\mu p_\mu - m) \psi = 0, \eqno(1)
$$
where $\gamma^\mu$- are Dirac matrices, $p_\mu$ -are energy and
impulse of the particle, $m$ - is a particle mass and $\psi$ is
the fermion wave function.
\par
If to introduce the following projecting operators
$$
P_L = \frac{(1 - \gamma^5)}{2} = \left(\begin{array}{cc} 0 & 0 \\
0 & 1 \end{array}\right), \quad P_R = \frac{(1 + \gamma^5)}{2} = \left(\begin{array}{cc} 1 & 0 \\
0 & 0 \end{array}\right), \eqno(2)
$$
which form the full system
$$
P_L^2 = P_L, \quad P_R^2 = P_R, \quad P_L + P_R =1, \quad P_L P_R
= 0, \eqno(3)
$$
then we can write $\psi$ in the form
$$
\psi = \psi_R + \psi_L, \eqno(4)
$$
where
$$
\psi_R = P_R \psi  ,\quad \psi_L = P_L \psi,
$$
are the left and right spin projections.
\par
Above we have considered free fermions, then to consider their
productions, annihilations and et cetera, it is necessary to
introduce their interactions. For introducing interactions in the
framework of the quantum field theory \cite{11} we use derivative
lengthening and the principle of gauge invariance. For example, in
the electrodynamics:
$$
p_\mu \to p_\mu - i e A_\mu(x), \quad A'_\mu(x) = A_\mu(x) +
\alpha(x), \eqno(5)
$$
$$
\bar \psi' = \bar \psi \exp(i \alpha(x)),\quad \psi' = \psi \exp(i
\alpha(x)), \eqno(6)
$$
where $A_\mu$ is a vector field, $e$- a couple constant of
electromagnetic interactions, $\alpha$ - a phase of gauge
transformation. Then equation (1) is invariant relative to the
above gauge transformation. The Lagrangian of electromagnetic
interactions is
$$
L(\bar\psi, \psi, A_\mu) = -i e \bar\psi \gamma^\mu \psi A_\mu
\equiv -i e (\bar\psi_L + \bar\psi_R) \gamma^\mu (\psi_L + \psi_R)
A_\mu. \eqno(7)
$$
In analogous manner we also introduce interactions in the strong
interaction \cite{12}, \cite{13} and the electroweak model
\cite{1}. In electromagnetic and strong interactions the left and
right components of fermions participate in a symmetric manner and
$$
\psi = \psi_L + \psi_R,
$$
while left doublets and right singlets of fermions participate in
weak interactions \cite{1} (in charged current on weak
interactions only left components of fermions participate). The
Lagrangian of the interaction of the electron neutrino with the
electron-positron (charged leptons) by $W$ bosons, has the
following form:
$$
L(W) = - \frac{ ig_W}{ \sqrt{2}} [\bar \nu_{L e} \gamma^\mu e_L
W^{+}_\mu +  \bar e_L \gamma^\mu \nu_{L e} W^{-}_\mu]  . \eqno(8)
$$
For neutrinos it means that only the left components of neutrinos
and antineutrinos participate in the weak interactions.
\par
Experiments \cite{4spiral} were performed to measure the neutrino
spirality. The result is the following - neutrino has the left
spirality and antineutrino has the right spirality. These
characteristic of neutrino was used to formulate the electroweak
model \cite{1}.
\par
So, in electromagnetic and strong interactions the left and right
components of fermions (quarks and charged leptons) participate in
a symmetric form:
$$
\bar \psi = \bar \psi_L + \bar \psi_R, \quad \psi = \psi_L +
\psi_R, \eqno(9)
$$
while only the left components of quarks and leptons participate
in the weak interactions:
$$
\bar \psi \to \bar \psi_L + 0, \quad \psi \to \psi_L + 0,
$$
i. e. for neutrino
$$
\bar \nu \equiv \bar \nu_L, \quad \nu \equiv \nu_L. \eqno(10)
$$

\section{Majorana neutrino}

\par
In 1937 Majorana found an equation \cite{2maj} for a fermion with
spin $\frac{1}{2}$. Then it became clear that this fermion could
be only a neutral particle since the particle and antipartic\-le
are joined in one representation. The Majorana equation for
neutrino is \cite{2maj}
$$
\begin{array}{c}
i (\widehat{\sigma}^\mu d_\mu) \nu_R - m^M_R \epsilon \nu^{*}_R
=0, \\
 i(\widehat{\sigma}^\mu d_\mu) \nu_L - m^M_L \epsilon \nu^{*}_L
=0, \end{array} \eqno(11)
$$
where $\widehat{\sigma}^\mu \equiv (\sigma^0, {\rm {\sigma}})$,
$\sigma^\mu \equiv (\sigma^0, - {{\sigma}})$, ${{\sigma}}$ is
Pauli matrices,
$$
\epsilon = \left ( \begin{array}{cc} 0, 1 \\
-1, 0 \end{array} \right ).
$$
These equations describe two completely different neutrinos with
masses $m^M_R$ and $m^M_L$ which do not possess any additive
numbers and neutrinos are their own antineutrinos; i.\,e.,
particles differ from antiparticles only in spin projections. Now
it is possible to introduce the following two Majorana neutrino
states:
$$
\begin{array}{c} \chi_L = \nu_L + (\nu_L)^c, \\
 \chi_R = \nu_R + (\nu_R)^c.
 \end{array} \eqno(12)
$$
$$
(\nu_{L R})^c = C \bar \nu^T_{L R},
$$
where $C$ - is a matrix of charge conjugation, $T$ - is
transposition \cite{4}-\cite{6}. Formally the above Majorana
equation (11) can be rewritten in the following form:
$$
(\gamma^\mu \partial_\mu  + m) \chi(x) =0,  \eqno(13)
$$
with the Majorana condition ($\chi \equiv \chi_{L R}$)
$$
C \bar \chi^T (x) = \xi \chi (x) ,  \eqno(14)
$$
where $\xi$ is a phase factor ($\xi = \pm 1$)
\par
It is necessary to stress that
$$
\bar \chi (x) \gamma^\mu \chi (x) = 0;  \eqno(15)
$$
i.\,e., vector current of Majorana neutrino is equal to zero.
\par
In 1950s the Majorana neutrino study was very extensive
\cite{3pauli}. Later it stopped. By that time the spirality of
neutrinos had been measured \cite{4spiral}.
\par
From the above consideration now we know that in strong and
electromag\-netic interactions the left and right components of
fermions participate symmetri\-cally while in the weak
interactions only the left components of fermions participate (in
neural current the right components of fermions are also present,
but in a nonsymmetric form).
\par
What is a Majorana neutrino? As it has been stressed above, that
Majorana particle $\chi$ with spin $\frac{1}{2}$ has two spin $\pm
\frac{1}{2}$ projections. The (left) component with projection
$-\frac{1}{2}$ is correlated with neutrino $\nu_L$ and the (right)
component with projection $+\frac{1}{2}$ is correlated with
antineutrino $(\nu_L)^c$. Here is an analogy with the
electromagnetic or strong interactions (see exp. (7)), where the
left and right components of fermions participate symmetrically.
In order to produce the Majorana neutrino, we must have an
interaction where neutrino and antineutrino participate
symmetrically. From the experimental data we have known that in
the weak interactions neutrino and antineutrino are produced in
separate processes but not in one process symmetrically. So,
Majorana neutrino can be produced only in the interaction which
does not differentiate particle from antiparticle but
differentiates spin projections of the particles (fermions). By
analogy with electromagnetic interactions we can introduce
Majorana neutrino current in the following form:
$$
j^\mu =  \bar \chi (x) \gamma^\mu \chi (x),  \eqno(16)
$$
but as it is stressed in exp. (15) this value is zero. We can
introduce the gauge transformation for the following Majorana
neutrinos $\chi_L$ and $\chi_R$:
$$
\begin{array}{c} \chi_L' = \exp(-i \beta) \chi_L \\
\chi_R' = \exp(i \beta) \chi_R \end{array}, \eqno(17)
$$
then we can also introduce Majorana charge $g_M$. It is clear that
this charge will differ from Dirac charge $g_W$. If this Majorana
particle interacts with a Dirac particle it must have the Dirac
charge (i.e., Majorana neutrino must have a double charge). Since
usually it is supposed that masses of $\chi_L$ and $\chi_R$
neutrino differ very much then this gauge transformation must be
strongly violated.
\par
A reaction with the double beta decay with two electrons
$$
(Z, A) \rightarrow (Z+2, A) + e^{-}_1 + e^{-}_2 + \bar \nu_{e 1} +
\bar \nu_{e 2} , \eqno(18)
$$
is possible if $M_A (Z,A) > M_A (Z+2,A)$.
\par
If neutrino is a Majorana particle ($\chi_L = \nu_L + (\nu_L)^c$),
then the following neutrinoless double beta decay is possible:
$$
(Z, A) \rightarrow (Z+2, A) + e^{-}_1 + e^{-}_2 , \eqno(19)
$$
if $M_A (Z,A) > M_A (Z+2,A)$.
\par
The lepton part of the amplitude of the above two neutrino decay
has the following form \cite{5}, \cite{6}, \cite{8} :
$$
\bar e(x) \gamma_\rho \frac{1}{2}(1 \pm \gamma_5) \nu_j  \bar e(y)
\gamma_{\sigma} \frac{1}{2} (1 \pm \gamma_5)  \nu_k (y) .
\eqno(20)
$$
After substituting the Majorana neutrino propagator and its
integrating on the momentum of virtual neutrino, the lepton
amplitude gets the following form:
$$
-i \delta_{jk} \int \frac{d^4 q}{(2 \pi)^4} \frac{e^{- i q
(x-y)}}{q^2 - m_j^2} \bar e(x) \gamma_\rho \frac{1}{2}(1 \pm
\gamma_5) (q^\mu \gamma_\mu + m_j) \frac{1}{2} (1 \pm \gamma_5)
\gamma_{\sigma} e(y). \eqno(21)
$$
If we use the following expressions:
$$
\ \frac{1}{2}(1 - \gamma_5) (q^\mu \gamma_\mu + m_j) \frac{1}{2}
(1 - \gamma_5)  = m_j \frac{1}{2} (1 - \gamma_5),  \eqno(22)
$$
$$
 \frac{1}{2}(1 - \gamma_5) (q^\mu \gamma_\mu + m_j)
\frac{1}{2} (1 + \gamma_5)  = q^\mu \gamma_\mu \frac{1}{2} (1 +
\gamma_5),  \eqno(23)
$$
then we see that in the case when there are only left currents
(expression (22)) we get a deposit only from the neutrino mass
part,
$$
-i \delta_{jk} \int \frac{d^4 q}{(2 \pi)^4} \frac{e^{- i q
(x-y)}}{q^2 - m_j^2} \bar e(x) \gamma_\rho \frac{1}{2}(1 -
\gamma_5) (q^\mu \gamma_\mu + m_j) \frac{1}{2} (1 - \gamma_5)
\gamma_{\sigma} e(y) =
$$
$$
= -i \delta_{jk} \int \frac{d^4 q}{(2 \pi)^4} \frac{e^{- i q
(x-y)}}{q^2 - m_j^2} \bar e(x) \gamma_\rho m_j \frac{1}{2} (1 -
\gamma_5) \gamma_{\sigma} e(y), \eqno(24)
$$
while at the presence of the right currents (expression (23))
$$
-i \delta_{jk} \int \frac{d^4 q}{(2 \pi)^4} \frac{e^{- i q
(x-y)}}{q^2 - m_j^2} \bar e(x) \gamma_\rho \frac{1}{2}(1 -
\gamma_5) (q^\mu \gamma_\mu + m_j) \frac{1}{2} (1 + \gamma_5)
\gamma_{\sigma} e(y) =
$$
$$
= -i \delta_{jk} \int \frac{d^4 q}{(2 \pi)^4} \frac{e^{- i q
(x-y)}}{q^2 - m_j^2} \bar e(x) \gamma_\rho q^\mu \gamma_\mu
\frac{1}{2} (1 + \gamma_5) \gamma_{\sigma} e(y) , \eqno(25)
$$
the amplitude includes the term proportional to four momentum $q$
in the neutrino propagator.
\par
So, we see that if neutrino is a Majorana particle, then the
neutrinoless double decay will take place. As we have shown above,
the standard weak interactions can produce only Dirac neutrinos.
The Majorana neutrino can produce only the interaction which does
not differentiate neutrino from antineut\-rino and, then neutrino
and antineutrino are left and right spin projections of one
particle. Until now this interaction has not been discovered.
Detection of the neutrinoless double decay will be an indication
on the existence of such interaction, therefore experiments with
very high precision are needed. Consideration of the problem, in
what concrete interaction the Majorana neutrino can be produced,
will be continued in subsequent works.

\section{Conclusion}

In this work the next problem was considered: what type (Dirac or
Majorana) of neutrinos is produced in the standard weak
interactions? We have come to a conclusion that these interactions
can produce only Dirac neutrinos but not Majorana's. It means that
this neutrino will be produced in another type of interactions.
Namely, Majorana neutrinos will be produced in the interaction
which differentiates spin projections but cannot distinguish
neutrino (particle) from antineutrino (antiparticle). Such
interactions have not been discovered yet. Therefore experiments
with very high precision are needed to detect the neutrinoless
double decay.


\end{document}